\documentclass[aps,epjc,twocolumn,floatfix,superscriptaddress,nofootinbib,showpacs,showkeys,preprintnumbers]{revtex4}
\usepackage{graphicx}
\usepackage{dcolumn}  
\usepackage{bm}       
\usepackage{palatino}

\preprint{preprint number}
\usepackage{epsfig}
\begin{document}
\title{ Nuclear attenuation of high energy two-hadron system \\
in the string model}
\author{N.~Akopov}
\author{L.~Grigoryan} 
\author{Z.~Akopov, \\
Yerevan Physics Institute, Br.Alikhanian 2, 375036 Yerevan, Armenia } 
\begin {abstract}
\hspace*{1em} Nuclear attenuation of the two-hadron system is considered in 
the string model. The two-scale model and its improved version with two 
different choices of constituent formation time and sets of parameters 
obtained earlier for the single hadron attenuation, are used to describe 
available experimental data for the $z$-dependence of subleading hadron, 
whereas satisfactory agreement with the experimental data has been observed. A 
model prediction for $\nu$-dependence of the nuclear attenuation of 
the two-hadron system is also presented.
\end {abstract}
\maketitle
\section{Introduction}
\normalsize 
\hspace*{1em}  Nuclear Attenuation (NA) of high energy hadrons is a well known 
tool for investigation of an early stage of the hadronization process, which can not
be described in the framework of the existing theory of strong interactions
 (perturbative QCD), because of major role of "soft" interactions.
Nevertheless, 
there are many phenomenological models, which describe, rather qualitatively,  
existing experimental data for single hadron NA~\cite{A1}-\cite{A10}. Also 
some predictions for the attenuation of multi-hadron systems leptoproduced in 
nuclear matter in the framework of the string model~\cite{A11}-\cite{A12} were 
done. It was argued that measurements of NA of multi-hadron systems can remove 
some ambiguities in determination of the parameters describing strongly 
interacting systems at the early stage of particle production: formation time 
of hadrons and cross-section of the intermediate state interaction inside the 
nucleus. Recently, for the first time, data~\cite{A13} on the two-hadron system 
NA ratio measured as a function of relative energy of the subleading hadron has been 
published. Therefore, in this work we attempt to describe these data based on the Two-Scale
Model (TSM)~\cite{A4} and Improved Two-Scale Model (ITSM)~\cite{A10}.
We present also predictions for the $\nu$-dependence of two-hadron 
system NA within the same model. Also, possible mutual screening of 
hadrons in string and its experimental verification is discussed.
\section{Theoretical framework}
\begin{figure*}[!ht]
\center{
\includegraphics[width=10cm]{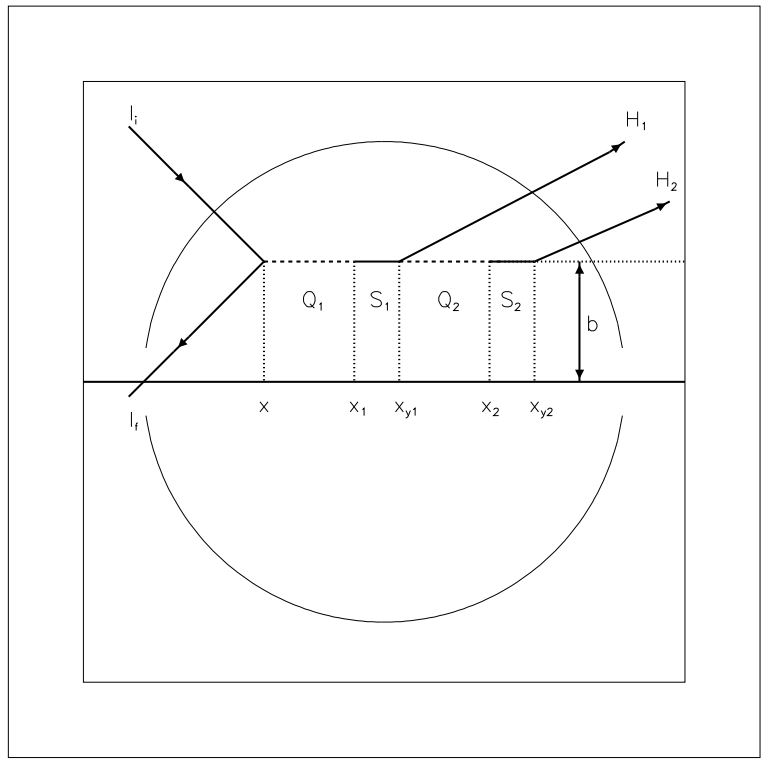}
\caption{{\it Leptoproduction of two-hadron system on nuclear target. 
Details see in text.}}
\label{fig:xx1}
}
\end{figure*}

\normalsize
\hspace*{1em}  In articles ~\cite{A11}-\cite{A12} the process of leptoproduction 
of two-hadron system on a nucleus with atomic mass number $A$ was  
theoretically considered for the first time:
\begin{eqnarray}
          {l_i + A \rightarrow l_f + h_1 + h_2 + X\hspace{0.15cm},}
\end{eqnarray}

where the hadrons $h_1$ and $h_2$ carry fractions $z_1$ and $z_2$ of the total
available energy. The NA ratio for that process can be expressed in form:
\begin{eqnarray}
{R_M^{2h} = 2d\sigma_A(\nu,Q^2,z_1,z_2)/Ad\sigma_D(\nu,Q^2,z_1,z_2)\hspace{0.15cm},} 
\end{eqnarray}
where $d\sigma_A$ and $d\sigma_D$ are the cross-sections for the reaction (1)
on nuclear and deuterium targets, respectively, $\nu$ and $Q^2$ denote the
energy of the virtual photon and square of its four-momentum. 
One can picture the reaction (1) as shown in Fig.~\ref{fig:xx1}. The
interaction of the 
lepton with nucleon occurs at the point $(b,x)$, where the intermediate state $q$ 
begins to propagate ($b$ and $x$ are impact parameter and longitudinal 
coordinate of DIS point). At some points the string breaks, and as a result 
the first constituents of hadrons $h_1$ and $h_2$ are produced at the points $(b,x_1)$ and
$(b,x_2)$. Also, at the points $(b,x_{y1})$ and 
$(b,x_{y2})$ the yo-yo of the hadrons $h_1$ and $h_2$ is formed
 (by a "yo-yo" system we assume, that the colorless system with valence
content and quantum numbers of the final hadron is formed, but without
its "sea" partons).

In the string model there are simple connections between the points
$x_{y1}-x_1=z_1L$ and $x_{y2}-x_2=z_2L$, where 
$L$ is the full hadronization length, $L = \nu/\kappa$,\hspace{0.15cm}$\kappa$
is the string tension (string constant).
The NA ratio can be presented in the following form:
\begin{eqnarray}
{R_M^{2h}\approx\frac{1}{2} \int{d^2b} \int_{-\infty}^{\infty}dx
\int_{x}^{\infty}dx_1\int_{x_1}^{\infty}{dx_2\rho(b,x)}}
\end{eqnarray}
\begin{eqnarray}
\nonumber
{[D(z_1,z_2,x_1-x,x_2-x)W_0(h_1,h_2;b,x,x_1,x_2)+}
\end{eqnarray}
\begin{eqnarray}
\nonumber
{D(z_2,z_1,x_1-x,x_2-x)W_0(h_2,h_1;b,x,x_1,x_2)]\hspace{0.15cm},} 
\end{eqnarray}
where $D(z_1,z_2,l_1,l_2)$ (with $l_1 < l_2$) is the distribution of the 
formation lengths $l_1$ and $l_2$ of the hadrons and $\rho(b,x)$ is the nuclear
density function normalized to unity. $W_0$ is the probability that neither
the hadrons $h_1$, $h_2$ nor the intermediate state leading to their 
production (initial and open strings) interact inelastically in the nuclear 
matter: 
\begin{eqnarray}
{W_0(h_1,h_2;b,x,x_1,x_2) =(1-Q_1-S_1-}
\end{eqnarray}
\begin{eqnarray}
\nonumber
{(H_1+Q_2+S_2+H_2-H_1(Q_2+S_2+H_2)))^{(A-1)}\hspace{0.15cm},}       
\end{eqnarray}
where $Q_1$ and $Q_2$ are the probabilities for the initial string to be  
absorbed in the nucleus within the intervals $(x,x_1)$ and $(x_{y1},x_2)$,
respectively.
$S_i\hspace{0.15cm}(i=1,2)$ is the probability for the open string containing 
the first constituent parton for $h_i$ to be
absorbed in the nucleus within an interval $(x_i,x_{yi})$, and 
$H_i\hspace{0.15cm}(i=1,2)$ is the probability for the $h_i$ to interact
inelastically in the nuclear matter, starting from the point $x_{yi}$.
\begin{figure*}[!ht]
\center{
\includegraphics[width=12cm]{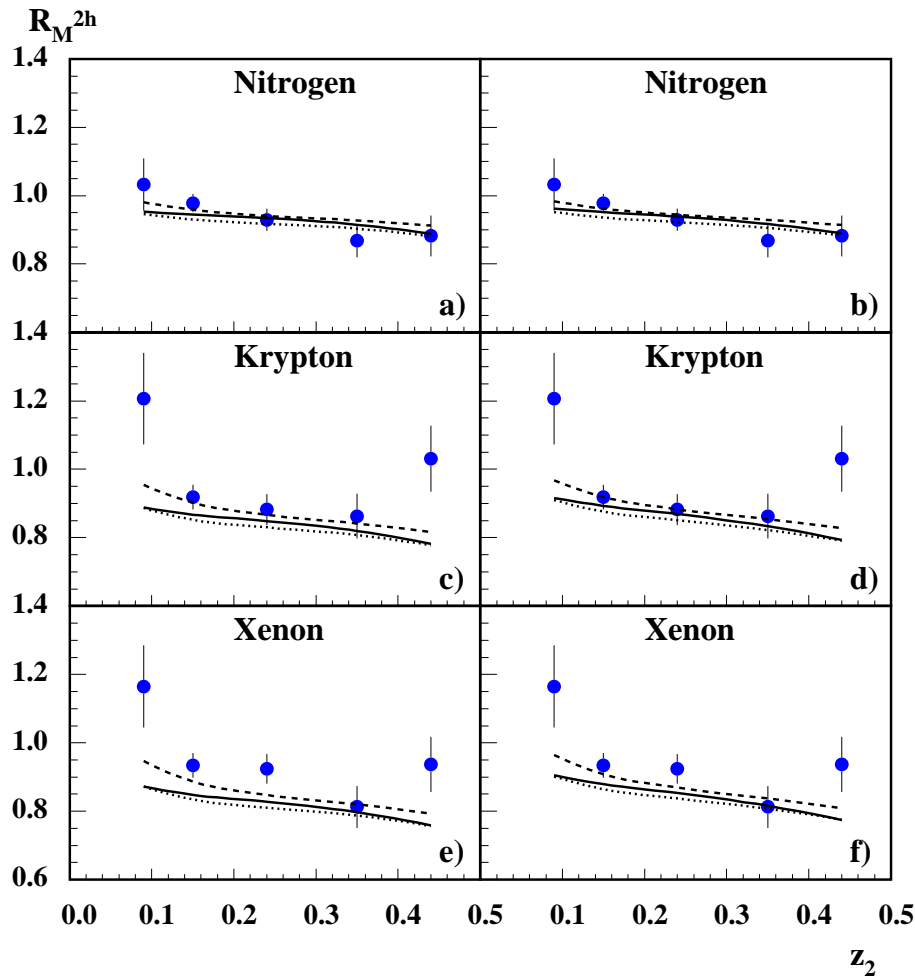}
\caption{{\it  Double ratio $R_M^{2h}$ as a function of $z_2$.  
The explanations for the points and curves are written in the text.}}
\label{fig:xx2}
}
\end{figure*}
\begin{figure*}[!ht]
\center{
\includegraphics[width=12cm]{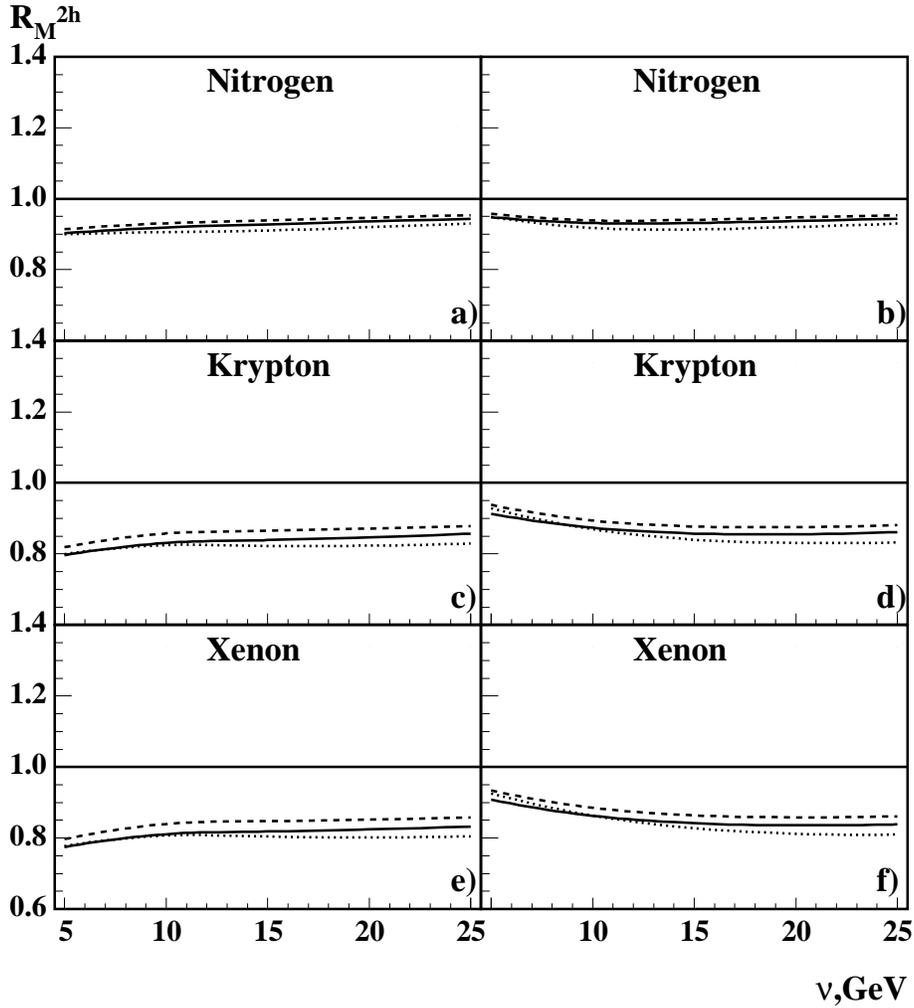}
\caption{{\it  Double ratio $R_M^{2h}$ as a function of $\nu$.  
The explanations for the curves are written in the text.}}
\label{fig:xx3}
}
\end{figure*}
\begin{figure*}[!ht]
\center{
\includegraphics[width=10cm]{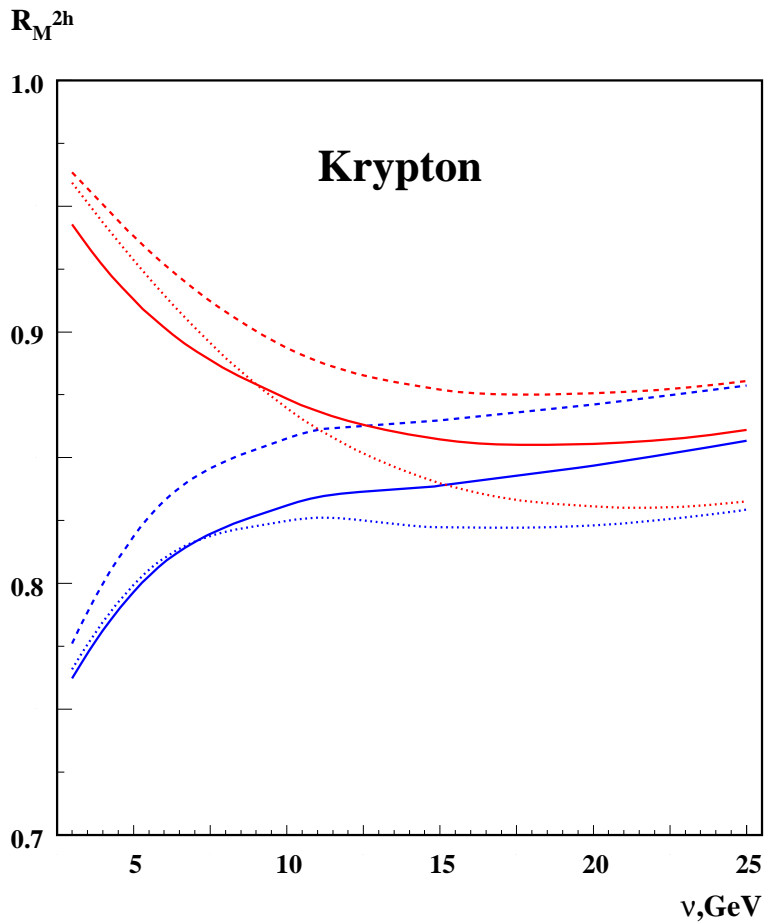}
\caption{{\it  Double ratio $R_M^{2h}$ as a function of $\nu$.  
The explanations for the curves are written in the text. 
Lower curves correspond to the case that two hadrons attenuates 
independently (full attenuation) and upper curves correspond to the case that 
only first produced hadron attenuates (partial attenuation)}}
\label{fig:xx4}
}
\end{figure*}
\begin{figure*}[!ht]
\center{
\includegraphics[width=10cm]{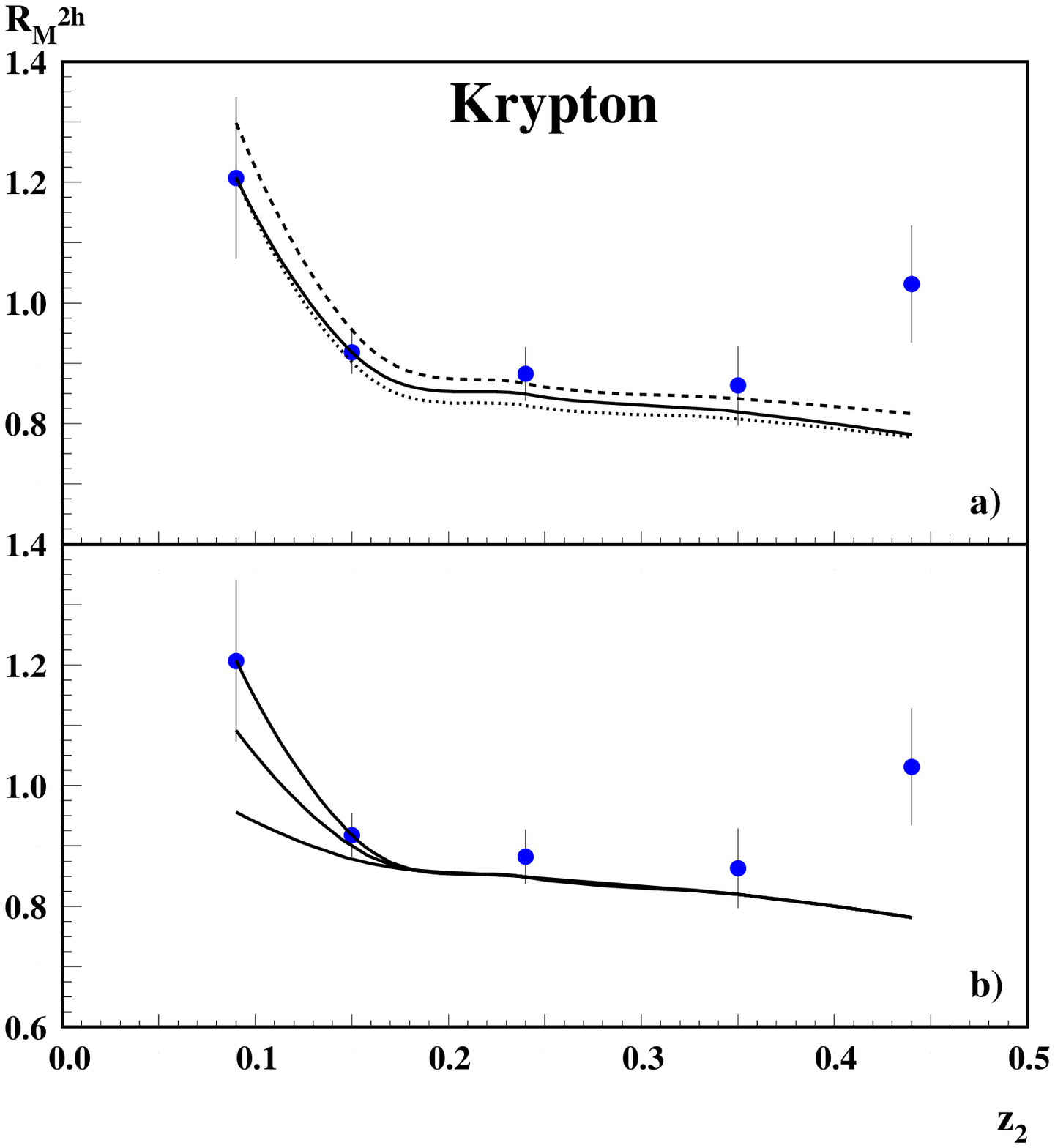}
\caption{{\it Double ratio $R_M^{2h}$ as a function of $z_2$. The result of
the proton inclusion in model is demonstrated (see for details Appendix 1.)   
The explanations for the points and curves are written in the text. 
}}
\label{fig:xx5}
}
\end{figure*}
The probabilities $Q_1, Q_2, S_1, S_2, H_1, H_2$ can be calculated
using the general formulae:
\begin{eqnarray}
 {P(x_{min},x_{max})=\int_{x_{min}}^{x_{max}}\sigma_P\rho(b,x)dx\hspace{0.15cm},}  
\end{eqnarray}
where the subscript $P$ denotes the particle (initial string, open string
or hadron), $\sigma_P$ its inelastic cross section on nucleon target, and
$x_{min}$ and $x_{max}$ are the end points of its path in the x direction, as
is shown in Fig.~\ref{fig:xx1}.

\section{Experimental situation}
 Recently the HERMES Collaboration has obtained, for the first time, data on double 
hadron attenuation~\cite{A13}. The following double ratio for leading and 
subleading hadrons has been measured:
\begin{eqnarray}
{R_M^{2h}(z_2) = (d^2N(z_1,z_2)/dN(z_1))_A/}
\end{eqnarray}
\begin{eqnarray}
\nonumber
{(d^2N(z_1,z_2)/dN(z_1))_D\hspace{0.15cm},}  
\end{eqnarray}
where $z_i = E_i/\nu$, $E_i$ is the energy of i-th hadron, $A$ and $D$ marks 
that interaction takes place on nuclear and deuterium targets, $d^2N(z_1,z_2)$ 
is the number of events with at least two hadrons (leading and 
subleading hadrons in one event). According to the experimental conditions leading 
hadron must have $z_1 >$ 0.5, and $dN(z_1)$ is the number of events with
at least one hadron with $z_1 >$ 0.5. In this experiment double ratio
$R_M^{2h}(z_2)$ was considered as a function of $z_{subleading} = z_2$. 
Two sets of experimental data are presented:
\begin{enumerate}
\item the leading-subleading combinations $++, --, +0$, $0+, -0, 0-, 00$ only;
\item all hadron pairs except those with invariant mass near $\rho^0$.
\end{enumerate}
In our work, we will use the first set, because, in our opinion, it contains less contribution
from hadrons produced in diffractive and diffractive dissociation processes. 
The relevant region for the investigation of the two-hadron system NA ratio as a
function of $z_2$ is the one between $z_2$ = 0.1 and $z_2$ = 0.4. At $z_2 <$ 0.1 
the contributions from the slow hadrons coming from the target fragmentation 
become large. At $z_2 >$ 0.4 it is difficult to distiguish 
leading and subleading hadron because in this region 
$z_1$\hspace{0.15cm}$\approx$\hspace{0.15cm}$z_2$\hspace{0.15cm}$\approx$\hspace{0.15cm}0.5.

Let us also consider the following question - to what extent are these data free from
contribution of diffractive $\rho^0$ mesons? 
For events $dN(z_1)$ corresponding corrections were made by experimentalists.
Here we discuss situation with $d^2N(z_1,z_2)$ only.
At a first glance the diffractive $\rho^0$ mesons do not give contribution in these 
events because of special conditions at choice of pairs of hadrons. They ensure 
that both hadrons in pair can not be produced from decay of a one $\rho^0$ 
meson, but do not forbid that one of the hadrons in the event was 
produced as a result of breaking of a diffractive $\rho^0$-meson. Then, another hadron
could be produced after the $\rho^0$-meson final state interaction.

\section{Results and Discussion}

 We have performed calculations for $R_M^{2h}(z_2)$ in the framework of TSM and its 
improved version. The ratio $R_M^{2h}$ is a function of three variables $z_1, z_2$ 
and $\nu$, but for comparison with the available experimental data we present 
$R_M^{2h}$ as a function of $z_2$ only, after integration over other variables
according to the experimental conditions 
$(0.5 < z_1 < 1 - z_2, 7 < \nu < 23.5 GeV, p_h > 1.4GeV/c)$.
In this paper we also present model predictions for $R_M^{2h}$ as a function 
of $\nu$, while performing integrations over $z_1, z_2$ in the corresponding kinematic 
regions. They are chosen to be close to the conditions of the available experimental data
$(0.5 < z_1 < 0.9, 0.1 < z_2 < 1 - z_1)$. The upper limit of $z_1 < 0.9$ relates to the case 
of two hadrons in
final state, as well as for the single hadron this limit has to be equal to unity.
All theoretical curves presented here were obtained with an assumption
that the final hadrons are pions. We used two expressions for Constituent Formation 
Time (CFT). As a first expression for CFT the equation (4.1) from~\cite{A12} 
(marked as CFT1) was used. Second expression was taken in a form according to
Ref.~\cite{A14}, which means, that as CFT for the first produced constituent 
quark of the first hadron we use $\tau_{c1} = (1 - z_1 - z_2)L$, and for the first 
produced constituent quark of the second hadron  - $\tau_{c2} = (1 - z_2)L$ (marked as CFT2). 
Three sets of parameters (including nuclear density functions)
corresponding to the minimum values of $\chi^2$ from Tables 1, 2, 3 
Ref.~\cite{A10} were used for calculations. The value of string tension was 
fixed at  $\kappa$ = 1~GeV/fm. Results for the double ratio $R_M^{2h}$ as a 
function of $z_2$ are presented on Fig.~\ref{fig:xx2}. On panels a), c),
e)  
three options of the theoretical curves are shown: solid curves correspond to the 
TSM with CFT1; dashed curves correspond to the ITSM with CFT2; dotted curves 
correspond to the ITSM with CFT1.
According to the ideology of the string model, the transverse size of the string is 
much less than longitudinal one. It means that the hadrons produced from the string have close impact
parameters, and could partly screen one another, which in turn must lead to the weakness of 
NA (partial attenuation).
To study this effect and to compare with 
the basic supposition, that two hadrons attenuate independently (full attenuation), 
we consider partial attenuation in it's extreme case, when two hadrons fully screen 
one another, and as a result the two-hadron system attenuates as a single hadron. 
The results of calculations within this conditions are shown in panels b), d), f).
The two-hadron system will attenuate as a single hadron also when the two
final hadrons appear as a result of breaking of one of the resonance. For instance, 
combinations $+0(-0)$ and $0+(0-)$ can be obtained as products of decay of
the $\rho^+$($\rho^-$) mesons produced via fragmentation mechanism in the nucleus, when 
the decay occured outside of the
nucleus.
Comparison with the experimental data for $z_2$-dependence shows that
difference between versions is smaller than the experimental errors,
consequently, different versions of model can not be distinguished by
means of comparison with these data. Calculations with full and partial
NA give close results. Theoretical curves quite satisfactory describe data
for nitrogen. In case of krypton and xenon targets the situation is more ambiguous. While
three middle points are described satisfactory, two extreme points
corresponding to lower and higher values of $z_2$ are in much worse agreement. 
In our opinion, that possible reason could be that our model does not contain necessary
ingredients for quantitative description of these points.\\
Let us briefly discuss mechanisms which, we believe, give considerable
contribution in the extreme points discussed above, but are not included in 
present model. 
First mechanism which can lead to the increasing of the NA ratio at
lower value of $z_2$ is that the part of subleading hadrons in nucleus
are protons, which are produced in abundance at small $z$, and in this region 
they have value of NA ratio larger than unity. We will try to estimate the 
contribution of protons in Appendix 1.
Second mechanism which can lead to the increasing of the NA ratio at
lower $z_2$ is the rescattering of produced hadrons in the nucleus, in result 
hadrons spend part of theirs energy for production of slow hadrons.
Consequently, more slow hadrons arise in nucleus, than in deuterium, and, 
despite absorption, the multiplicity
ratio in this region can become close to and even larger than unity. 
Our model does not take into
account the final state interactions of the produced hadrons, consequently, 
at present we can not calculate or estimate contribution of this mechanism.
Concerning second extreme point at higher value of $z_2$, which is equal
$z_2 = 0.44$,  we suppose that the double hadron attenuation ratio in this point is on the
order of unity because there are two more additional mechanisms, which are not included in
present model. 
The first one had to do with the pairs of pions appearing in a result of breaking
of coherently produced diffractive $\omega$-mesons, for which the coherent cross section
depends proportional to the atomic mass number as $~A^2$. 
As a result, the NA ratio for heavy nuclei raises.
It is very difficult to estimate the contribution of this mechanism 
without implementation of additional free parameters.
Second mechanism is connected with the smallness of integration region over $z_1$,
which in case of $z_2 = 0.44$ is equal to 0.06 \footnote{this value of 0.06 is
in fact related to the case of only two hadrons in final state, more hadrons in final state
lead to the decreasing of the integration range}. The NA ratio for two-hadron system is
proportional to the ratio of integrals over $z_1$ on nucleus and deuterium. Taking into
account Fermi motion or nucleon-nucleon correlations can lead to the extension of the
integration region in nucleus and in a result to the increasing of NA
ratio (see details in Appendix 2).
 The model gives close results for the two-hadron system NA ratio for the krypton
 and xenon targets.   

Fig.~\ref{fig:xx3} shows the prediction of model for $\nu$-dependence of
double ratio $R_M^{2h}$ for
nitrogen, krypton and xenon targets. 
On the panels a), c), e) three varieties of the 
theoretical curves  are shown: solid curves correspond to the TSM
with CFT1; dashed curves correspond to the ITSM with CFT2; dotted curves
correspond to the ITSM with CFT1. On panels b), d), f) are shown the same 
curves calculated with additional condition that only first produced hadron 
attenuates (partial attenuation). Easy to see, that curves corresponding to 
the full and partial attenuation have different behavior at low values of 
$\nu$. In Fig.~\ref{fig:xx4} it is shown, as example, case of krypton
only. Curves marked 
as in Fig.~\ref{fig:xx3}. Lower curves correspond to the case of full
attenuation and 
upper curves correspond to the case that only first produced hadron 
attenuates (partial attenuation). The measurement of NA ratio in the region 
of $\nu$ from 3$GeV$ to 10$GeV$ is allowed to verify a supposition about
possible mutual screening of hadrons in string. We think that such 
experiment can be useful for comparison with the results obtained at RHIC 
by STAR Collaboration~\cite{A15}, which state that two hadrons from one jet
absorbed more weakly than two hadrons from away-side jets.

\section{Conclusions}
\begin{itemize}
\item { String model~\cite{A11,A12} gives natural and simple mechanism for 
description of the two-hadron system NA, which allows to describe the available experimental data
for $z$-dependence of subleading hadron on a satisfactory level, using the sets of parameters
obtained in Ref.~\cite{A10} for single hadron NA.}
\item {Comparison with the experimental data for $z_2$-dependence show that
difference between versions of model is smaller than experimental errors,
consequently, they can not be distinguished by means of comparison with 
these data.}
\item {Double ratio considered as a function of partial energy of
subleading hadron $z_2$ has weak sensitivity to the mutual 
screening of hadrons}
\item {It is of certain interest to also study other aspects of the 
two-hadron system production in nuclear medium. In particular we propose to measure
the $\nu$-dependence of NA, because, as we have shown, it is more sensitive to 
the mutual screening of hadrons than the $z_2$-dependence.
Investigation of the $\nu$-dependence in the region of $\nu$ from 
3$GeV$ to 10$GeV$ will allow better understanding of questions connected
with possible mutual screening of hadrons in the string. Corresponding
measurements can be performed at HERMES and JLab.}
\item {Estimations show that agreement with the experimental data can be
improved by means of inclusion of additional mechanisms which were not included
in the model we have presented.}
\end{itemize}

As a last remark concerning the description of data for two-hadron system
attenuation in other models.

There are at least two theoretical works which attempt to describe data for
two-hadron system NA: first of them based on BUU transport model~\cite{A16}, 
and second based on the so called energy loss model~\cite{A17}.
\begin{acknowledgments}
We would like to acknowledge P. Di Nezza, who has initiated the double 
hadron measurement at HERMES, as well as many other colleagues from the 
HERMES Collaboration for fruitful discussions.
\end{acknowledgments}
\begin{appendix}
\section*{APPENDIX 1}
In this appendix we will discuss considerable difference between our model 
and the experimental data at the point $z_2$=0.09 for krypton nucleus, and will 
try to understand the cause of this discrepancy. Our model, as mentioned 
above, takes into account pions only. But in heavy nuclei many slow protons are produced, in
addition to pions (in this discussion we do not distiquish kaons from pions). We want to show
that by including in consideration these protons one can improve agreement with data. Let us
present the two-hadron system NA ratio 
in the form:
\begin{eqnarray}
{R_M^{2h} = (1 - \alpha)R_M^{2\pi} + \alpha R_M^{\pi P}\hspace{0.15cm},} 
\end{eqnarray}
where $\alpha$ is the part of the events which contain pairs consisting from 
a fast pion and a slow proton, $R_M^{2\pi}$ and $R_M^{\pi P}$ are two-hadron 
system NA ratios in case when a pair consist of two pions and pion-proton, 
respectively. It is convenient to introduce a parameter 
$\beta = R_M^{\pi P}/R_M^{2\pi}$. Then $R_M^{2h}$ can be rewritten in the form:
\begin{eqnarray}
{R_M^{2h} = (1 + \alpha(\beta - 1))R_M^{2\pi}\hspace{0.15cm}.}
\end{eqnarray}
Parameter $\beta$ can be defined from the data on single hadron attenuation, 
if one can assume that the fast and the slow hadrons in the event are produced
independently (correlation between them can be neglected). Then
$\beta\approx R_M^{P}/R_M^{\pi}$, where $R_M^{P}$ and
$R_M^{\pi}$ are single hadron NA ratios for proton and pion respectively. 
In addition to the $z_2 = 0.09$ point we make estimates for the next point $z_2 = 0.15$ also,
because contribution of protons in this point can still be considerable. Using available 
data for krypton~\cite{A18}, we have extrapolated it in the point $z = 0.09 (0.15)$ 
and obtained $R_M^{P}\approx1.5 (1.425)$ and $R_M^{\pi}\approx0.85 (0.85)$, 
which gave $\beta\approx1.765 (1.676)$. Then we determined the parameter $\alpha$
from eq. (8). For $z_2 = 0.09 (0.15)$ the model gives $R_M^{2\pi}\approx0.89 (0.87)$, 
while the experimental value is $R_M^{2h}\approx1.2 (0.92)$ which gives
$\alpha\approx0.47 (0.087)$. The results of calculations with the model improved by means of 
inclusion of the protons, as was discussed above, are shown in Fig. 5a for
the krypton case. Three theoretical 
curves are shown: the solid curve corresponds to the TSM with CFT1; dashed curve 
corresponds to the ITSM with CFT2; dotted curve corresponds to the ITSM with CFT1. 
Such a value of $\alpha$ for the first point seems to be too big. But one can note that still, 
this correction works in the right direction and improves agreement of theory with 
the experimental data even if real value of parameter $\alpha$ is smaller than 
that following from our estimate. In Fig. 5b the dependence on value 
of parameter $\alpha$ at $z_2 = 0.09$ is shown. As an example we take the case of TSM with
CFT1 calculated for three values of $\alpha$ equal to $0.47$ (upper curve), $0.30$ 
(middle curve), $0.10$ (lower curve). Simultaneously we proportionally changed 
values of $\alpha$ in the next experimental point also. The estimates with other
versions of the model give close results.

\section*{APPENDIX 2}
As mentioned in text, the Fermi motion of nucleons in nucleus and the
presence of nucleon-nucleon correlations (fluctons) can improve
agreement with the experimental data at $z_2 = 0.44$ by means of extending 
the range of integration.\\
Let us first consider the influence of the Fermi motion on increase of
$z_{max}$. We would like to remind the reader, that in case of scattering on rest nucleon
$z_{max}=1$. The total centre of mass energy of the secondary hadrons $W$, and in
case of taking into account Fermi momentum we mark it as $W_F$. Then, after
averaging over the angle between virtual photon and nucleon momenta, it can be
presented as:
\begin{eqnarray}
{W_{F}^2 = M^2 -Q^2 + 2E_N\nu = M^2 -Q^2 + 2M\nu_{eff}\hspace{0.15cm},}
\end{eqnarray}
where $M$ and $E_N$ are mass and energy of nucleon respectively,
$\nu_{eff}$ is the value of $\nu$ which give on the rest nucleon value of
$W$ equal $W_F$. Then for $z_{max}$ we obtain:
\begin{eqnarray}
{z_{max} = \frac{\nu_{eff}}{\nu} = \frac{E_N}{M}\hspace{0.15cm}.}
\end{eqnarray}
Now we can estimate the influence of the Fermi momentum. If we
take the average Fermi momentum for middle and heavy nucleus to be equal to 
0.25~$GeV/c$~\cite{fermi}, then:
$z_{max}=$1.035 and NA ratio with taking into account that also the
integration region for single hadrons  will be increased (about 10\%),
$R_M^{2h}$ at $z_2 = 0.44$ must be multiplied on factor 1.42. 

As an alternative mechanism the nucleon-nucleon correlations could be
considered in the framework of the flucton mechanism. That means, that a virtual
photon scatters on fluctons with the masses $M$, 2$M$, 3$M$, etc. If one takes into 
account only one- and two-nucleon correlations, one obtains:
\begin{eqnarray}
{z_{max} = \frac{\nu_{eff}}{\nu}\approx 1 + \alpha_{fl} \hspace{0.15cm},}
\end{eqnarray}
where $\alpha_{fl}$ is the probability of scattering on the two-nucleon
flucton. If, for the sake of an estimate, one takes $\alpha_{fl} = 0.01$ then
$z_{max}=1.01$ and the NA ratio
$R_M^{2h}$ at $z_2 = 0.44$ must be multiplied by a factor of 1.17.
We see that both mechanisms give considerable improvement for the last $z_2$ point.
\end{appendix}

\end{document}